\documentclass[prd,preprint,aps,showpacs]{revtex4}
\usepackage{graphics}
\newcommand{\newc}{\newcommand}
\newc{\gsim}{\lower.7ex\hbox{$\;\stackrel{\textstyle>}{\sim}\;$}}
\newc{\lsim}{\lower.7ex\hbox{$\;\stackrel{\textstyle<}{\sim}\;$}}
\newc{\gev}{\,{\rm GeV}}
\newc{\mev}{\,{\rm MeV}}
\newc{\ev}{\,{\rm eV}}

\newc{\kev}{\,{\rm keV}}
\newc{\tev}{\,{\rm TeV}}
\newc{\mz}{m_Z}
\newc{\mpl}{M_{Pl}}
\newc{\chifc}{\chi_{{}_{\!F\!C}}}
\newc\order{{\cal O}}
\newc\CO{\order}
\newc\CL{{\cal L}}
\newc\CY{{\cal Y}}
\newc\CH{{\cal H}}
\newc\CM{{\cal M}}
\newc\CF{{\cal F}}
\newc\CD{{\cal D}}
\newc\CN{{\cal N}}
\newc{\eps}{\epsilon}
\newc{\re}{\mbox{Re}\,}
\newc{\im}{\mbox{Im}\,}
\newc{\invpb}{\,\mbox{pb}^{-1}}
\newc{\invfb}{\,\mbox{fb}^{-1}}
\newc{\yddiag}{{\bf D}}
\newc{\yddiagd}{{\bf D^\dagger}}
\newc{\yudiag}{{\bf U}}
\newc{\yudiagd}{{\bf U^\dagger}}
\newc{\yd}{{\bf Y_D}}
\newc{\ydd}{{\bf Y_D^\dagger}}
\newc{\yu}{{\bf Y_U}}
\newc{\yud}{{\bf Y_U^\dagger}}
\newc{\ckm}{{\bf V}}
\newc{\ckmd}{{\bf V^\dagger}}
\newc{\ckmz}{{\bf V^0}}
\newc{\ckmzd}{{\bf V^{0\dagger}}}
\newc{\X}{{\bf X}}
\newc{\bbbar}{B^0-\bar B^0}

\newc{\sgn}{\mbox{sgn}\,}
\newc{\m}{{\bf m}}
\newc{\msusy}{M_{\rm SUSY}}
\newc{\munif}{M_{\rm unif}}
\newc{\slepton}{{\tilde\ell}}
\newc{\Slepton}{{\tilde L}}
\newc{\sneutrino}{{\tilde\nu}}
\newc{\selectron}{{\tilde e}}
\newc{\stau}{{\tilde\tau}}
\relax
\def\beq{\begin{equation}}
\def\eeq{\end{equation}}
\def\bea{\begin{eqnarray}}
\def\eea{\end{eqnarray}}
\newc{\ie}{{\it i.e.}}          \newc{\etal}{{\it et al.}}
\newc{\eg}{{\it e.g.}}          \newc{\etc}{{\it etc.}}
\newc{\cf}{{\it c.f.}}
\def\Dsl{\,\raise.15ex\hbox{/}\mkern-13.5mu D} 
\def\delsl{\raise.15ex\hbox{/}\kern-.57em\partial}
\def\Ksl{\hbox{/\kern-.6000em\rm K}}
\def\Asl{\hbox{/\kern-.6500em \rm A}}
\def\Qsl{\hbox{/\kern-.6000em\rm Q}}
\def\gradsl{\hbox{/\kern-.6500em$\nabla$}}
\def\bar#1{\overline{#1}}

\begin{document}
\title{Electroweak-Higgs Unification in the Two Higgs Doublet Model:
Masses and Couplings of the Neutral and Charged Higgs Bosons}
\author{J.L. D\'{\i}az-Cruz$^{(a)}$, and A. Rosado$^{(b)}$}
\affiliation{$^{(a)}$ Facultad de Ciencias F\'{\i}sico-Matem\'aticas, BUAP.\\
Apdo. Postal 1364, C.P. 72000 Puebla, Pue., M\'exico\\
$^{(b)}$ Instituto de F\'{\i}sica, BUAP. Apdo. Postal J-48, C.P.
72570 Puebla, Pue., M\'exico}
\date{\today}

\begin{abstract}
We obtain the mass spectrum and the Higgs self-coupling of the two
Higgs doublet model (THDM) in an alternative unification scenario
where the parameters of the Higgs potential $\lambda_i$
($i=1,2,3,4,5$) are determined by imposing their unification with
the electroweak gauge couplings. An attractive feature of this
scenario is the possibility to determine the Higgs boson masses by
evolving the $\lambda_i$,s from the electroweak-Higgs unification
scale $M_{GH}$ down to the electroweak scale. The unification
condition for the gauge ($g_1,g_2$) and Higgs couplings is written
as $g_1=g_2=f(\lambda_i)$, where $g_1=k_Y^{1/2} g_Y$, and $k_Y$
being the normalization constant. Two variants for the unification
condition are discussed; Scenario I is defined through the linear
relation: $g_1=g_2=k_H(i)\lambda_i(M_{GH})$, while Scenario II
assumes a quadratic relation: $g^2_1=g^2_2=k_H(i)\lambda_i(M_{GH})$.
In Scenario I, by fixing {\it ad hoc} $-k_H(5)=\frac{1}{2}
k_H(4)=\frac{3}{2} k_H(3)=k_H(2)=k_H(1) =1$, taking $\tan\beta=1$
and using the standard normalization ($k_Y=5/3$), we obtain the
following spectrum for the Higgs boson masses: $m_{h^0} = 109.1$
GeV, $m_{H^0} = 123.2$ GeV, $m_{A^0} = 115.5$ GeV, and $m_{H^{\pm}}
= 80.3$ GeV, with similar results for other normalizations such as
$k_Y=3/2$ and $k_Y=7/4$.
\end{abstract}

\maketitle

\setcounter{footnote}{0}
\setcounter{page}{2}
\setcounter{section}{0}
\setcounter{subsection}{0}
\setcounter{subsubsection}{0}


\section{Introduction}

The Standard Model (SM) of the strong and electroweak (EW)
interactions has met with extraordinary success; it has been already
tested at the level of quantum corrections
\cite{radcorrs1,radcorrs2}. These corrections give some hints about
the nature of the Higgs sector, pointing towards the existence of a
relatively light Higgs boson, with a mass of the order of the EW
scale,  $m_{\phi_{SM}} \simeq v$ \cite{hixjenser}.
 However, it is widely believed that the SM cannot be the final theory
of particle physics, in particular because the Higgs sector suffers
from naturalness problems, and we do not really have a clear
understanding of electroweak symmetry breaking (EWSB).

These problems in the Higgs sector can be stated as our present
inability to find a satisfactory answer to some questions regarding
its structure, which can be stated as follows:
\begin{enumerate}
\item What fixes  the size (and sign) of the  {\it dimensionful parameter}
$\mu^2_0$ that appears in the Higgs potential?. This parameter
determines the scale of EWSB in the SM; in principle it could be as
high as the Planck mass, however, it needs to be fixed to much lower
values.
\item What is the nature of  the quartic Higgs coupling $\lambda$?.
This parameter is not associated with a known symmetry, and we expect
all interactions in nature to be associated somehow with gauge forces,
as these are the ones we understand better \cite{myghyunif}.
\end{enumerate}

An improvement on our understanding of EWSB is provided by the
supersymmetric (SUSY) extensions of the SM \cite{softsusy}, where
loop corrections to the tree-level parameter  $\mu^2_0$ are under
control, thus making the Higgs sector more natural. The quartic
Higgs couplings is nicely related with gauge couplings through
relations of the form: $\lambda=\frac{1}{8} (g^2_2+g^2_Y)$. In the
SUSY alternative it is even possible to (indirectly) explain the
sign of  $\mu^2_0$ as a result loop effects and the breaking of the
symmetry between bosons and fermions. Further progress to understand
the SM structure is achieved in Grand Unified Theories (GUT), where
the strong and electroweak gauge interactions are unified at a
high-energy  scale ($M_{GUT}$) \cite{gutrev}. However, certain
consequences of the GUT idea seem to indicate that this unification,
by itself, may be too drastic (within the minimal SU(5) GUT model
one actually gets inexact unification, too large proton decay,
doublet-triplet problem, incorrect fermion mass relations, etc.),
and some additional theoretical tool is needed to overcome these
difficulties. Again, SUSY offers an amelioration of these problems.
When SUSY is combined with the GUT program, one gets a more precise
gauge coupling unification and some aspects of proton decay and
fermion masses are under better control
\cite{myradferm1,myradferm2}.

In order to verify the realization of SUSY-GUT in nature, it will be
necessary to observe plenty of new phenomena such as superpartners,
proton decay or rare decay modes.

As nice as these ideas may appear, it seems worthwhile  to consider
other approaches for physics beyond the SM. For instance, it has
been shown that additional progress towards understanding the SM
origin, can be achieved by postulating the existence of extra
dimensions. These theories have received much attention, mainly
because of the possibility they offer to address the problems of the
SM from a new geometrical perspective. These range from a new
approach to the hierarchy problem
\cite{ADD1,ADD2,ADD3,ADD4,RanSundrum} up to a possible explanation
of flavor hierarchies in terms of field localization along the extra
dimensions \cite{nimashmaltz}. Models with extra dimensions have
been applied to neutrino physics
\cite{abdeletal1,abdeletal2,abdeletal3,abdeletal4,Ioannisian:1999sw},
Higgs phenomenology \cite{myhixXD1,myhixXD2}, among many others. In
the particular GUT context, it has been shown that it is possible to
find viable solutions to the doublet-triplet problem
\cite{hallnomu1,hallnomu2}. More recently, new methods in strong
interactions have also been used as an attempt to revive the old
models (TC, ETC, topcolor, etc) \cite{fathix}. Other ideas have
motivated new types of models as well (little Higgs
\cite{littlehix}, AdS/CFT composite Higgs models \cite{hixdual},
etc).

In  this paper we are interested in exploring further an alternative
unification scenario, of weakly-interacting type, that could offer
direct  understanding of the Higgs sector too and was first
discussed in Ref.\cite{Aranda:2005st}. Namely, we shall explore the
consequences of a scenario where the electroweak $SU(2)_L\times
U(1)_Y$ gauge interactions  are unified with the Higgs
self-interactions at an intermediate scale $M_{GH}$. Here, we
explore further this idea within the context of the THDM, which
allows us to predict the Higgs spectrum of this model. The
dependence of our results on the choice for the normalization for
the hypercharge is also discussed, as well as possible test of this
EW-Higgs unification idea at future colliders, such as ILC. Besides
predicting the Higgs spectrum, namely the masses for the neutral
CP-even states ($h^0,H^0$), the neutral CP-odd state ($A^0$) and the
Charged Higgs ($H^\pm$), we also discuss the Higgs couplings to
gauge bosons and fermions. As we mentioned in our previous paper
\cite{Aranda:2005st}, it is relevant to compare our approach with
the so called Gauge-Higgs unification program, as they share some
similarities. We think that our approach is more model independent,
as we first explore the consequences of a parametric unification,
without really choosing a definite model at higher energies. In
fact, at higher energies both the SUSY models as well as the
framework of extra dimensions could work as ultraviolet completion
of our approach. The SUSY models could work because they allow to
relate the scalar quartic couplings to the gauge couplings, thanks
to the D-terms \cite{myghyunif}. On the other hand, within the
extra-dimensions it is also possible to obtain similar relations,
when the Higgs fields are identified as the extra-dimensional
components of gauge fields
\cite{XDGHix1,XDGHix2,ABQuiros1,ABQuiros2,ABQuiros3,Hosotani1,Hosotani2,
Hosotani3,Hosotani4,Hosotani5,Hosotani6,Hosotani7,Hosotani8}.
Actually, we feel that the work of
Ref.\cite{gauhixyuku1,gauhixyuku2} has a similar spirit to ours, in
their case they look for gauge unification of the Higgs
self-couplings that appear in the superpotential of the NMSSM, and
then they justify their work with a concrete model in 7D. However,
in the present work, we do not discuss further the unification of
the EW-Higgs couplings with the strong constant, which can be
realized within the context of extra-dimensional Gauge-Higgs unified
theories.

\begin{figure}[floatfix]
\begin{center}
\includegraphics{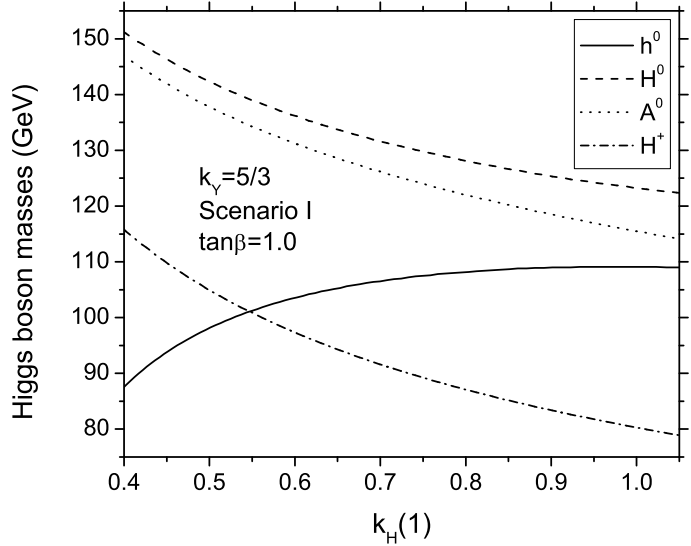}
\caption{Prediction for the Higgs boson masses as a function of
$k_H(1)$ in the context of the THDM with $k_Y=5/3$, in the frame of
Scenario I, taking $\tan\beta=1$ and $m_{top}=170.0$ GeV.}
\label{figure1}
\end{center}
\end{figure}

\begin{figure}[floatfix]
\begin{center}
\includegraphics{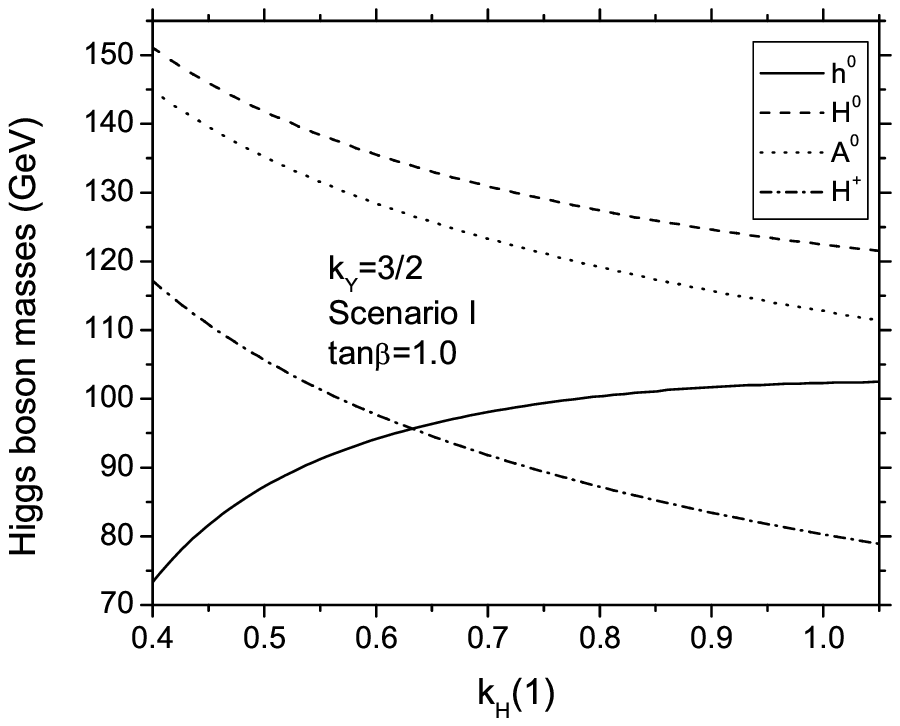}
\caption{Prediction for the Higgs boson masses as a function of
$k_H(1)$ in the context of the THDM with $k_Y=3/2$, in the frame of
Scenario I, taking $\tan\beta=1$ and $m_{top}=170.0$ GeV.}
\label{figure2}
\end{center}
\end{figure}

\begin{figure}[floatfix]
\begin{center}
\includegraphics{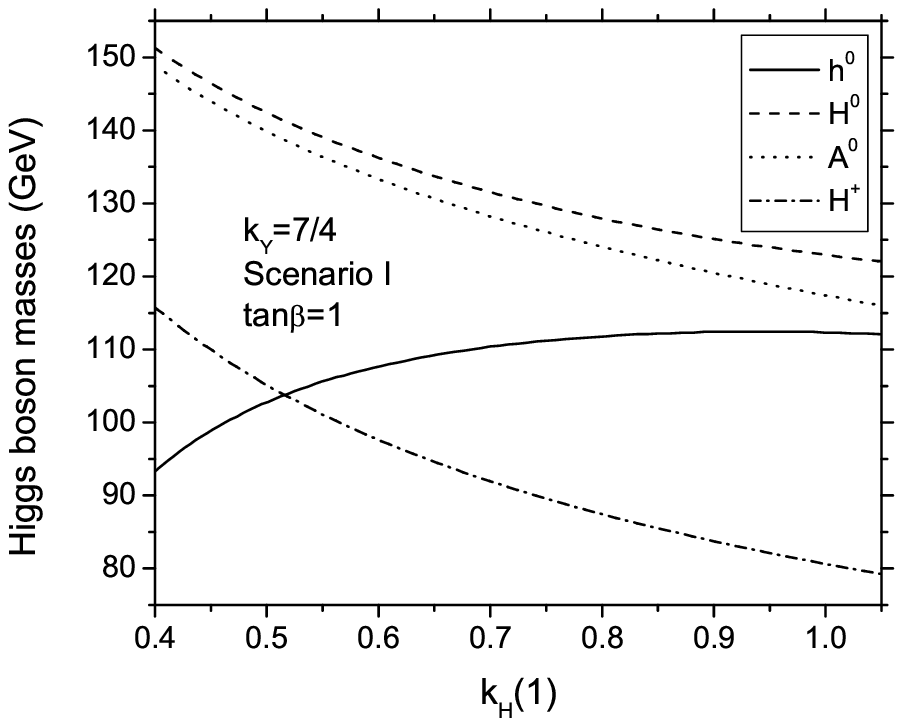}
\caption{Prediction for the Higgs boson masses as a function of
$k_H(1)$ in the context of the THDM with $k_Y=7/4$, in the frame of
Scenario I, taking $\tan\beta=1$ and $m_{top}=170.0$ GeV.}
\label{figure3}
\end{center}
\end{figure}

\begin{figure}[floatfix]
\begin{center}
\includegraphics{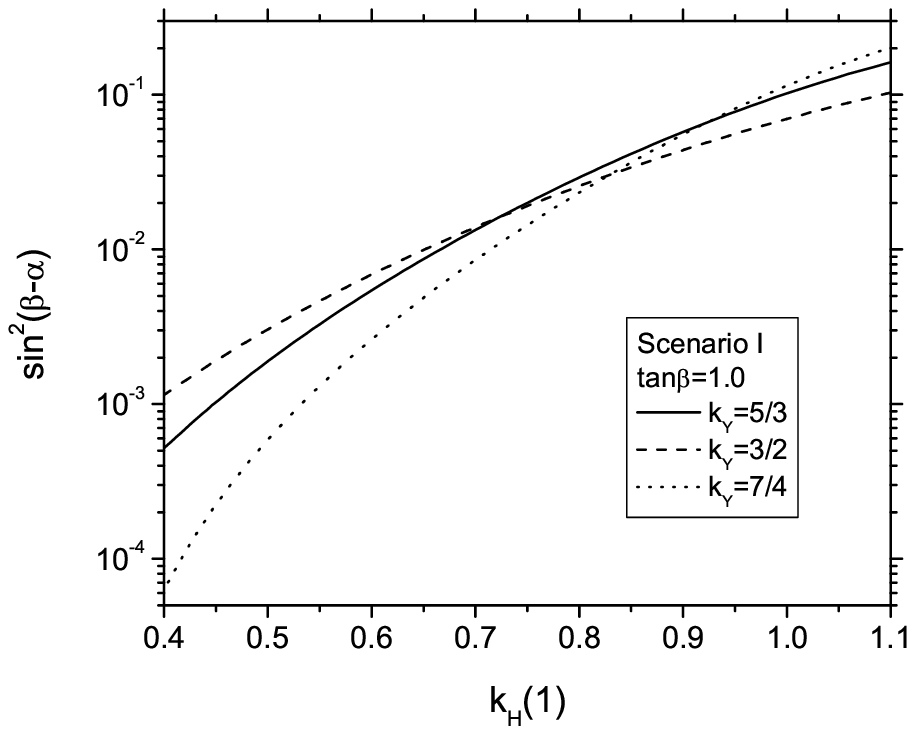}
\caption{Prediction for the $sin^2(\beta-\alpha)$ as a function of
$k_H(1)$ in the context of the THDM for $k_Y=5/3,3/2,7/4$, in the
frame of Scenario I, taking $\tan\beta=1$ and $m_{top}=170$ GeV.}
\label{figure4}
\end{center}
\end{figure}

\begin{figure}[floatfix]
\begin{center}
\includegraphics{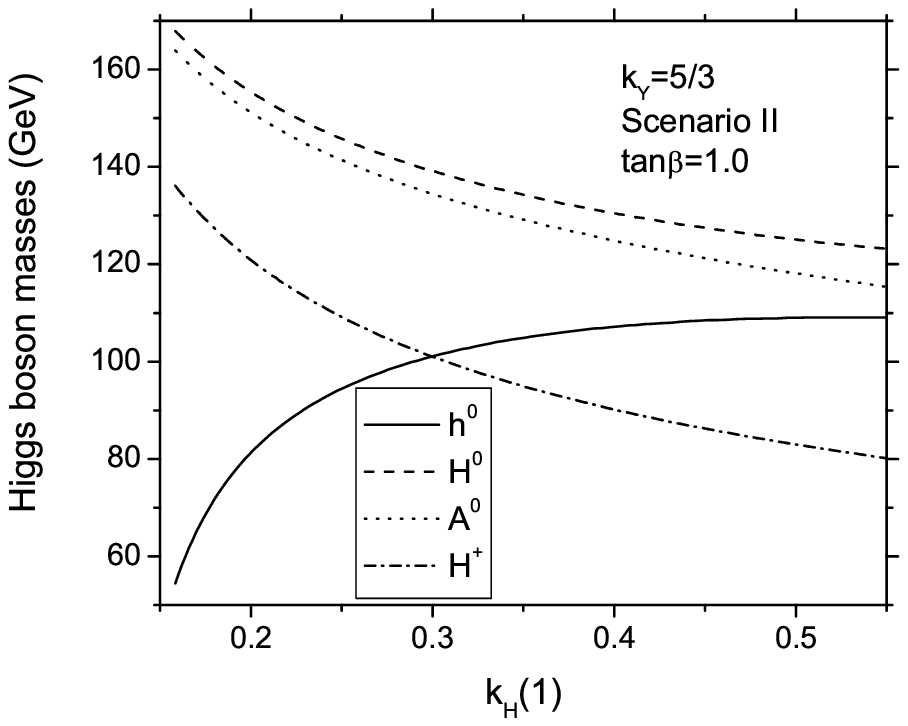}
\caption{Prediction for the Higgs boson masses as a function of
$k_H(1)$ in the context of the THDM with $k_Y=5/3$, in the frame of
Scenario II, taking $\tan\beta=1$ and $m_{top}=170.0$ GeV.}
\label{figure5}
\end{center}
\end{figure}

\begin{figure}[floatfix]
\begin{center}
\includegraphics{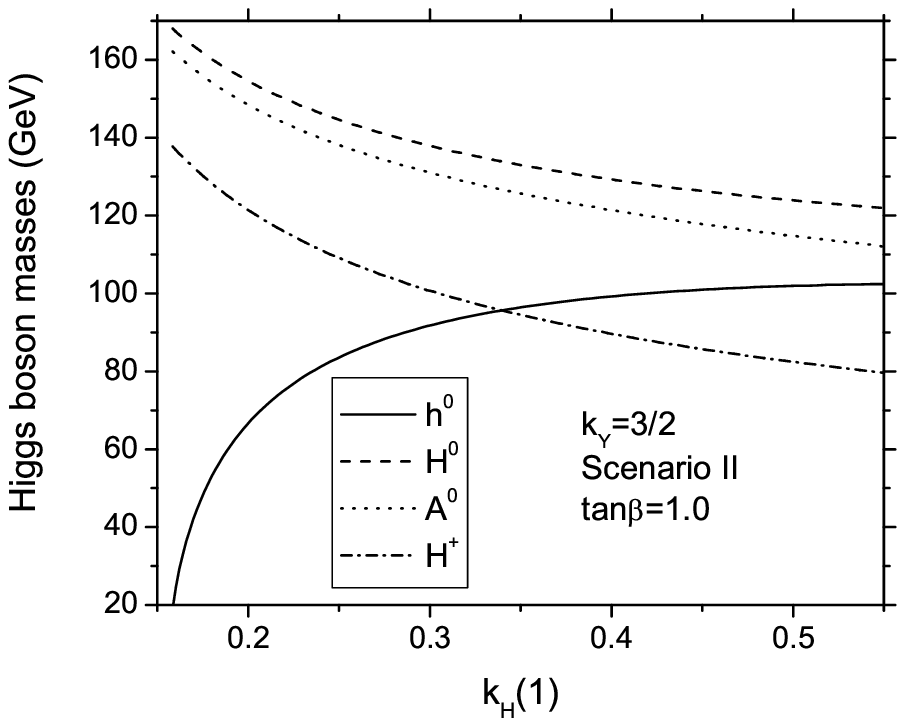}
\caption{Prediction for the Higgs boson masses as a function of
$k_H(1)$ in the context of the THDM with $k_Y=3/2$, in the frame of
Scenario II, taking $\tan\beta=1$ and $m_{top}=170.0$ GeV.}
\label{figure6}
\end{center}
\end{figure}

\begin{figure}[floatfix]
\begin{center}
\includegraphics{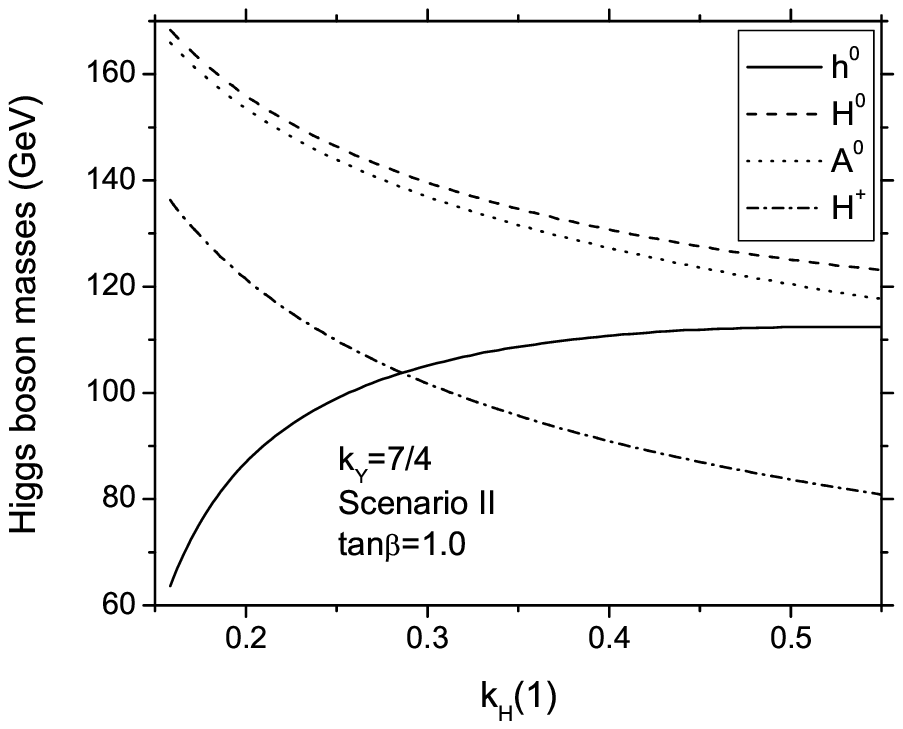}
\caption{Prediction for the Higgs boson masses as a function of
$k_H(1)$ in the context of the THDM with $k_Y=7/4$, in the frame of
Scenario II, taking $\tan\beta=1$ and $m_{top}=170.0$ GeV.}
\label{figure7}
\end{center}
\end{figure}

\begin{figure}[floatfix]
\begin{center}
\includegraphics{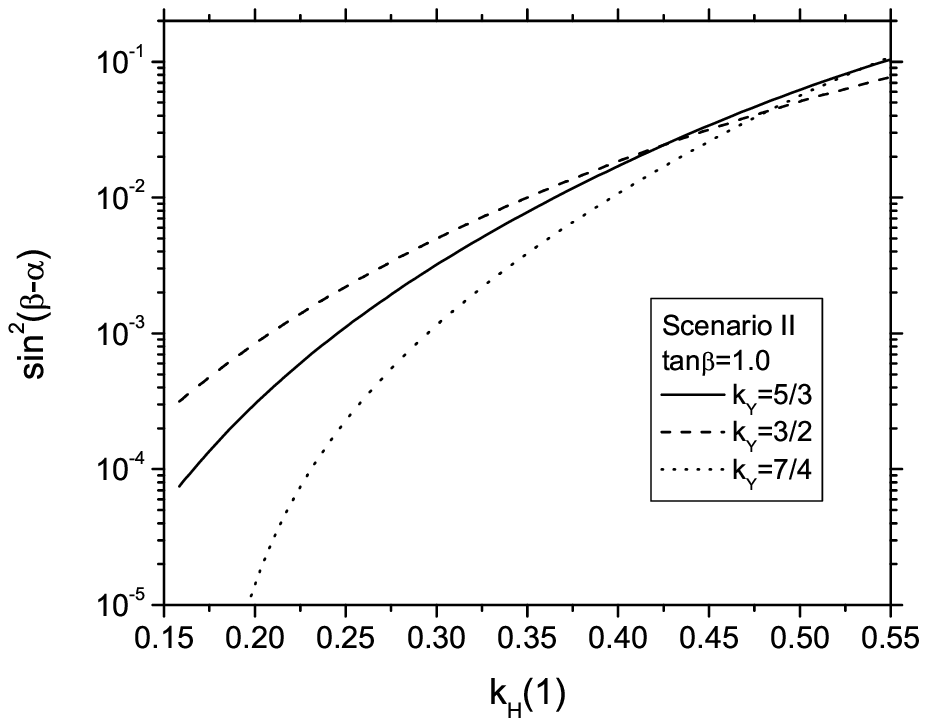}
\caption{Prediction for the $sin^2(\beta-\alpha)$ as a function of
$k_H(1)$ in the context of the THDM for $k_Y=5/3,3/2,7/4$, in the
frame of Scenario II, taking $\tan\beta=1$ and $m_{top}=170$ GeV.}
\label{figure8}
\end{center}
\end{figure}

\section{Gauge-Higgs unification in the SM: Review.}

In the EW-Higgs unified scenario, one assumes that there exists a
scale where the gauge couplings constants $g_1,g_2$, associated with
the gauge symmetry $SU(2)_L\times U(1)_Y$, are unified, and that at
this scale they also get unified with the SM Higgs self coupling
$\lambda$, i.e. $g_1=g_2=f(\lambda)$ at $M_{GH}$. The precise
relation between $g_1$ and $g_Y$ (the SM hypercharge coupling)
involves a normalization factor $k_Y$, i.e. $g_1=k_Y^{1/2}g_Y $,
which depends on the unification model. The standard normalization
gives $k_Y=5/3$, which is associated with minimal models such as
$SU(5), SO(10), E_6$. However, in the context of string theory it is
possible to have such standard normalization without even having a
unification group. For other unification groups that involve
additional $U(1)$ factors, one would have exotic normalizations too,
and similarly for the case of GUT models in extra-dimensions. In
what follows we shall present results for the cases: $k_Y=5/3,3/2$
and $7/4$, which indeed arise in string-inspired models
\cite{Dienes:1996yh}. Note that these values fall in the range $3/2
< 5/3 < 7/4$ and so they can illustrate what happens when one
chooses a value below or above the standard normalization. The form
of the unification condition will depend on the particular
realization of this scenario, which could be as generic as possible.
 However, in order to be able to make predictions for
the Higgs boson mass, we shall consider two specific realizations.
Scenario I will be based on the linear relation:  $g_2=g_1=k_H
\lambda(M_{GH})$, where the factor $k_H$ is included in order to
retain some generality, for instance to take into account possible
unknown group theoretical or normalization factors. Motivated by
specific models, such as SUSY itself, as well as an argument based
on the power counting of the beta coefficients in the RGE for scalar
couplings, {\it i.e.}, the fact that $\beta_\lambda$ goes as
$O(g^4)$, we shall also define Scenario II, through the quadratic
unification condition: $g^2_1=g^2_2=k_H\lambda (M_{GH})$. The
expressions for the SM renormalization group equations at the
two-loop level can be found in Ref.~\cite{langa1}.

In practice, one determines first the scale $M_{GH}$ at which $g_2$
and $g_1$ are unified, then one fixes the quartic Higgs coupling
$\lambda$ by imposing the unification condition and finally, by
evolving the quartic Higgs coupling down to the EW scale, we are
able to predict the Higgs boson  mass. For the numerical
calculations, discussed in Ref.\cite{Aranda:2005st}, we employed the
full two-loop SM renormalization group equations involving the gauge
coupling constants $g_{1,2,3}$, the Higgs self-coupling $\lambda$,
the top-quark Yukawa coupling $g_t$, and the parameter $k_Y$
\cite{rgebsm,langa1}. We also take the values for the coupling
constants as reported in the Review of Particle Properties
\cite{revPP}, while for the top quark mass we take the value
recently reported in~\cite{lasttopm1,lasttopm2}.

Now, let us summarize our previous results with the full numerical
analysis. For $k_Y=5/3$ we find that $M_{GH} \cong 1.0 \times
10^{13}$ GeV and by taking $\tan\beta=1$, results for the Higgs
boson mass are given as a function of the parameter $k_H$ over a
range $10^{-1} < k_H < 10^2$, which covers three orders of magnitude
(We stress here that the expected natural value for $k_H$ is 1). For
such a range of $k_H$, the Higgs boson mass takes the values: $176 <
m_H < 275$ GeV for Scenario I, while for $k_H=1$ we obtain a
prediction for the Higgs boson mass: $m_H=229,\,234,\,241$ GeV, for
a top quark mass of $m_{top}=165,\,170,\,175$ GeV
\cite{lasttopm1,lasttopm2}, respectively. On the other hand, for
Scenario II, we find that the Higgs boson mass can take the values:
$175 < m_H < 269$ GeV, while for $k_H=1$ we obtain:
$m_H=214,\,222,\,230$ GeV.

Then, when we compare our results with the Higgs boson mass obtained
from EW precision measurements, which imply $m_H \lsim 190$ GeV, we
notice that in order to get compatibility with such value, our model
seems to prefer high values of $k_H$. For instance, by taking the
lowest value that we consider here for the top mass, $m_t=165$ GeV,
and fixing $k_H=10^2$, we obtain the minimum value for the Higgs
boson mass equal to  $m_H=176$ GeV in Scenario I, while Scenario II
implies a minimal value that is slightly lower, $m_H=175$ GeV.

For $k_Y=3/2$ we find that $M_{GH}= 4.9 \times 10^{14}$ GeV, higher
than in the previous case, but for which one still gets a mass gap
between $M_{GH}$ and a possible $M_{GUT}$. In this case and by
taking $\tan\beta=1$ we find values a little higher for the Higgs
boson mass, for instance, for $k_H=1$, one gets
$m_H=225,\,232,\,238$ ($m_H=212,\,220,\,218$) GeV for scenario I
(II).

On the other hand, for $k_Y=7/4$ we find that $M_{GH}= 1.8 \times
10^{12}$ GeV, which is lower than that of the previous cases, and
has an even larger mass gap between $M_{GH}$ and a possible
$M_{GUT}$. In this case and by taking $\tan\beta=1$ we also find
values slightly higher for the Higgs boson mass, for instance, for
$k_H=1$, one gets $m_H=230,\,236,\,243$ ($m_H=215,\,223,\,231$) GeV
for scenario I (II).

At this point, rather than continuing discussions on the precise Higgs boson
mass, we would like to emphasize that our approach based on the EW-Higgs
unification idea is very  successful in giving a Higgs boson
mass that has indeed the correct order of magnitude, and that once
measured at the LHC we will be able to fix the
parameter $k_H$ and find connections with other approaches for
physics beyond the SM, such as the one to be discussed next.

In fact, for the Higgs boson mass range that is predicted in our
approach, it turns out that the Higgs will decay predominantly into
the mode $h\to ZZ$, which may provide us with good chances to
measure the Higgs boson mass within a precision of $5 \%$
\cite{Duhrssen:2004cv1,Duhrssen:2004cv2}, thus making it possible to
bound $k_H$ to within a few percent level. Further tests of our
EW-Higgs unification hypothesis would involve testing more
implications of the quartic Higgs coupling. For instance one could
use the production of Higgs pairs ($e^+ e^- \to \nu \bar{\nu} h h$)
at a future linear collider, such as the ILC. This is just another
example of the complementarity of future studies at LHC and ILC.

\begin{figure}[floatfix]
\begin{center}
\includegraphics{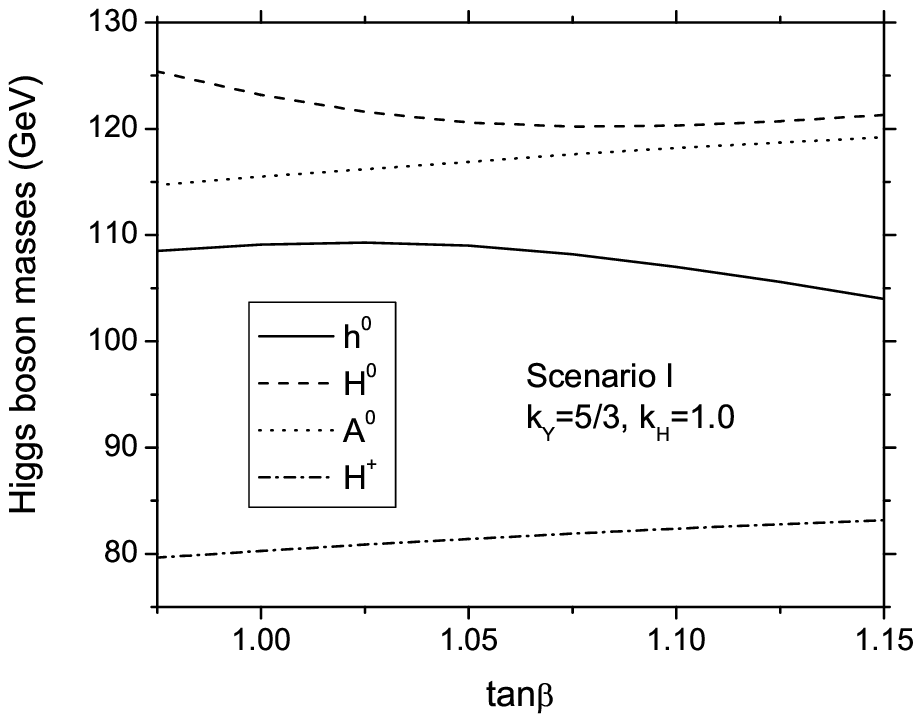}
\caption{Prediction for the Higgs boson masses as a function of
$\tan\beta$ in the context of the THDM with $k_Y=5/3$, in the frame
of Scenario I, taking $k_H(1)=1$ and $m_{top}=170.0$ GeV.}
\label{figure9}
\end{center}
\end{figure}

\begin{figure}[floatfix]
\begin{center}
\includegraphics{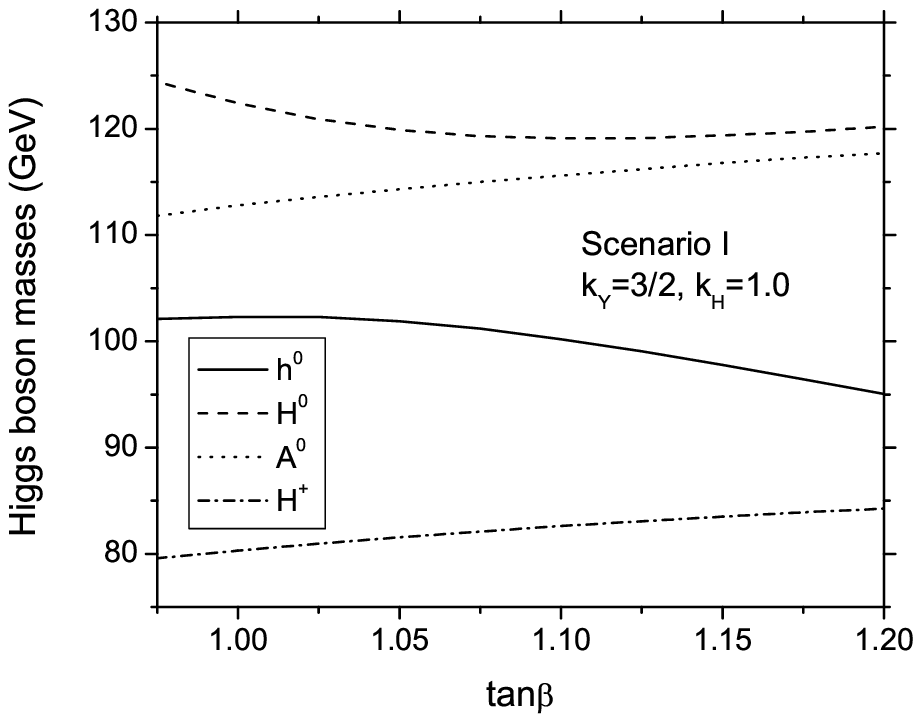}
\caption{Prediction for the Higgs boson masses as a function of
$\tan\beta$ in the context of the THDM with $k_Y=3/2$, in the frame
of Scenario I, taking $k_H(1)=1$ and $m_{top}=170.0$ GeV.}
\label{figure10}
\end{center}
\end{figure}

\begin{figure}[floatfix]
\begin{center}
\includegraphics{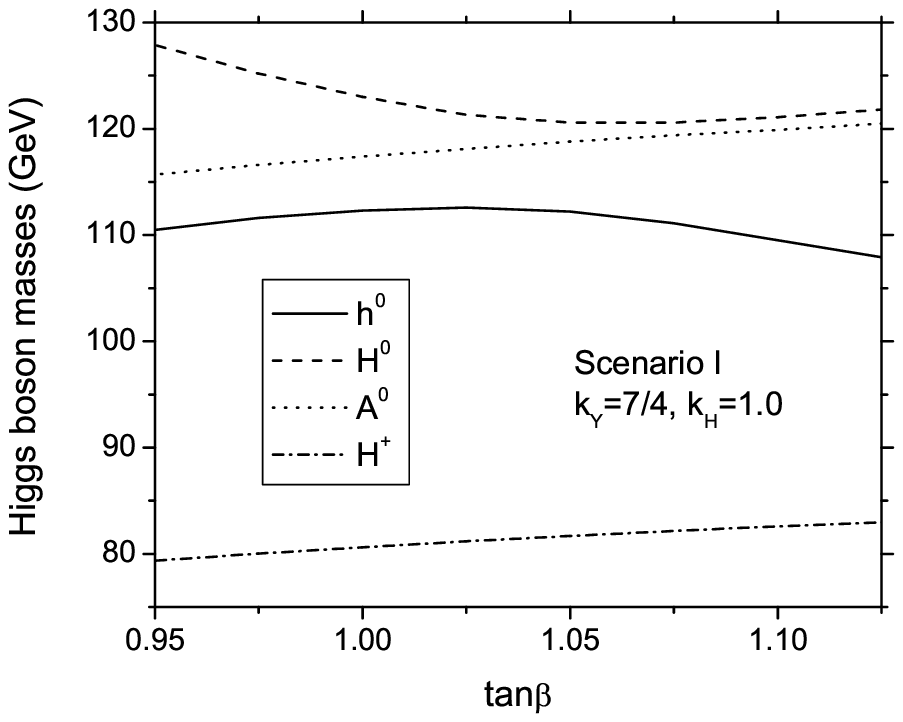}
\caption{Prediction for the Higgs boson masses as a function of
$\tan\beta$ in the context of the THDM with $k_Y=7/4$, in the frame
of Scenario I, taking $k_H(1)=1$ and $m_{top}=170.0$ GeV.}
\label{figure11}
\end{center}
\end{figure}

\begin{figure}[floatfix]
\begin{center}
\includegraphics{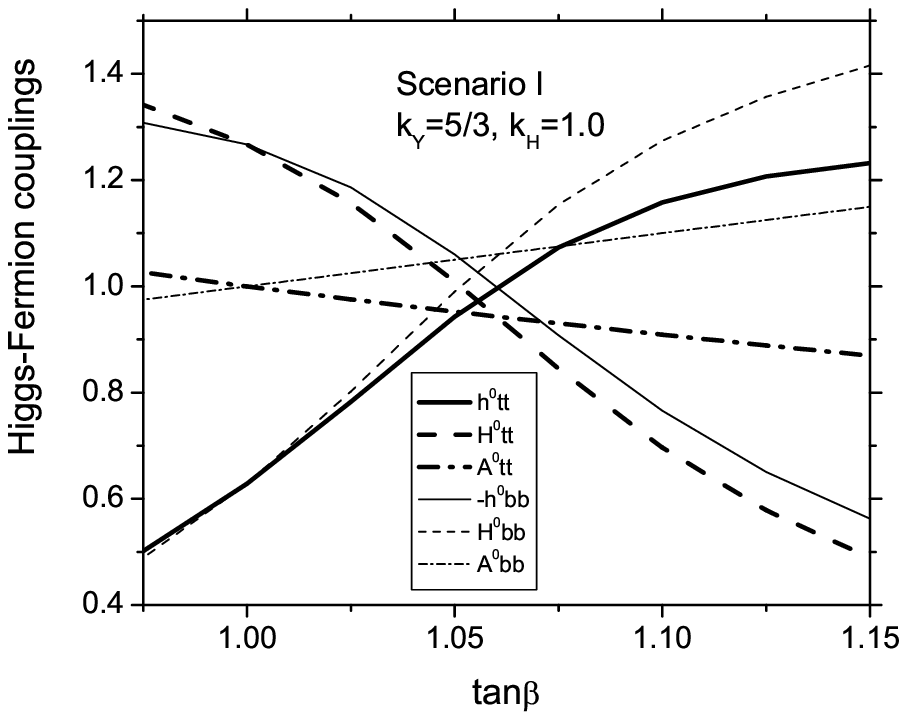}
\caption{Prediction for the Higgs-fermion couplings as a function of
$\tan\beta$ in the context of the THDM with $k_Y=5/3$, in the frame
of Scenario I, taking $k_H(1)=1$ and $m_{top}=170.0$ GeV. The curves
correspond to: 1) $h^0 t \bar{t}$, 2) $H^0 t \bar{t}$, 3) $A^0 t
\bar{t}$, 4) $-h^0 b \bar{b}$, 5) $H^0 b \bar{b}$ and 6) $A^0 b
\bar{b}$.} \label{figure12}
\end{center}
\end{figure}

\begin{figure}[floatfix]
\begin{center}
\includegraphics{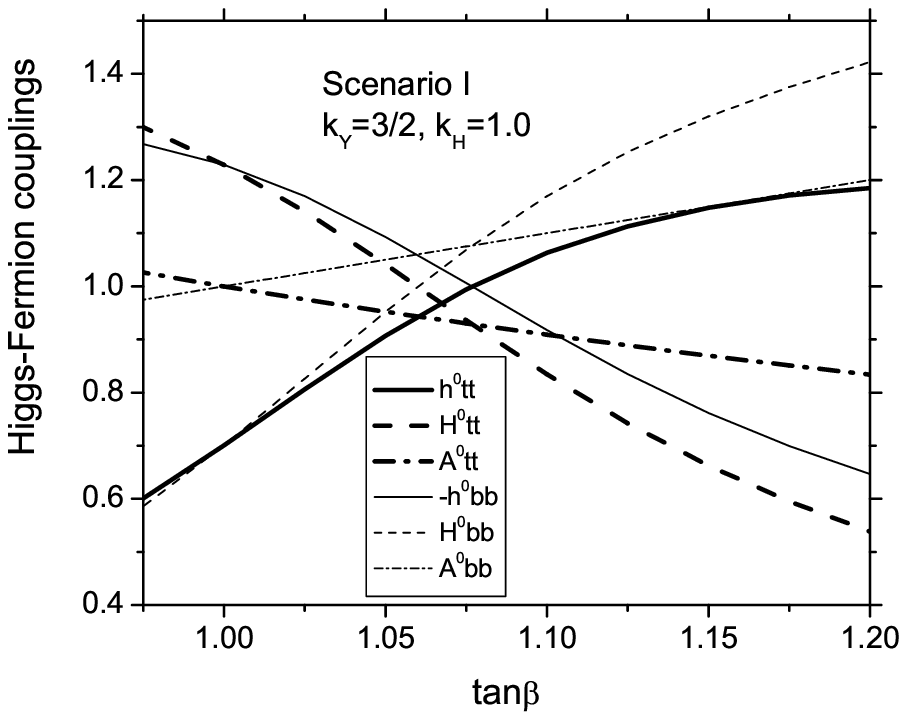}
\caption{Prediction for the Higgs-fermion couplings as a function of
$\tan\beta$ in the context of the THDM with $k_Y=3/2$, in the frame
of Scenario I, taking $k_H(1)=1$ and $m_{top}=170.0$ GeV. The curves
correspond to: 1) $h^0 t \bar{t}$, 2) $H^0 t \bar{t}$, 3) $A^0 t
\bar{t}$, 4) $-h^0 b \bar{b}$, 5) $H^0 b \bar{b}$ and 6) $A^0 b
\bar{b}$.} \label{figure13}
\end{center}
\end{figure}

\begin{figure}[floatfix]
\begin{center}
\includegraphics{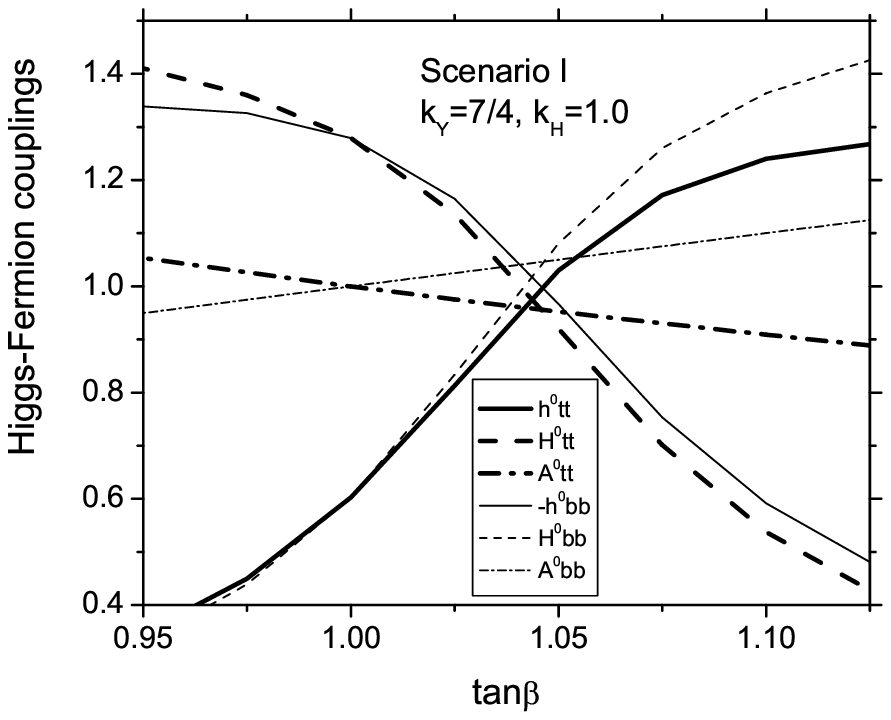}
\caption{Prediction for the Higgs-fermion couplings as a function of
$\tan\beta$ in the context of the THDM with $k_Y=7/4$, in the frame
of Scenario I, taking $k_H(1)=1$ and $m_{top}=170.0$ GeV. The curves
correspond to: 1) $h^0 t \bar{t}$, 2) $H^0 t \bar{t}$, 3) $A^0 t
\bar{t}$, 4) $-h^0 b \bar{b}$, 5) $H^0 b \bar{b}$ and 6) $A^0 b
\bar{b}$.} \label{figure14}
\end{center}
\end{figure}

\begin{figure}[floatfix]
\begin{center}
\includegraphics{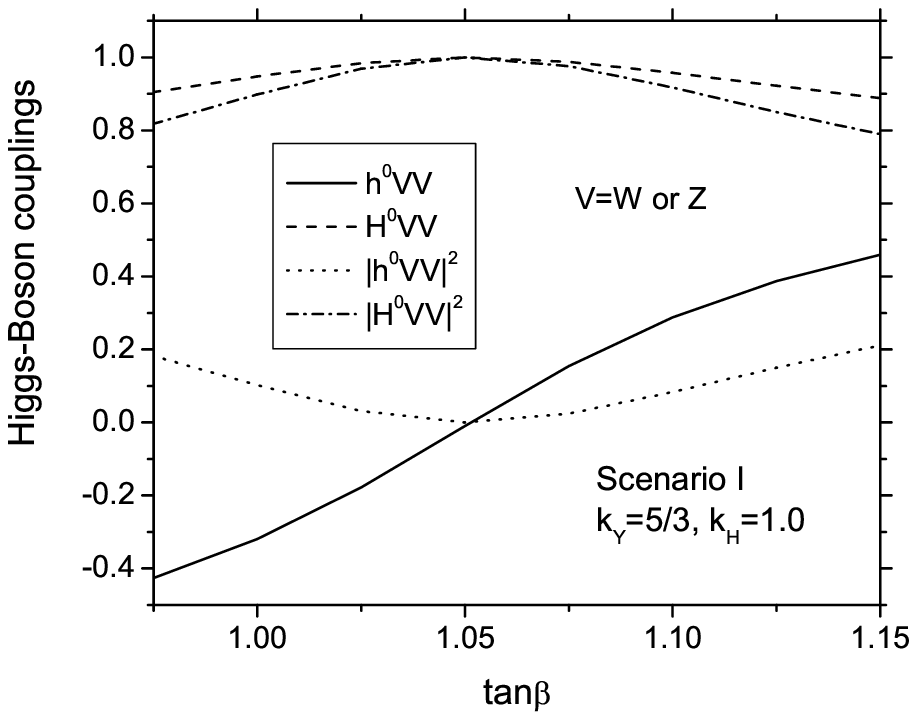}
\caption{Prediction for the Higgs-boson couplings as a function of
$\tan\beta$ in the context of the THDM with $k_Y=5/3$, in the frame
of Scenario I, taking $k_H(1)=1$ and $m_{top}=170.0$ GeV. The curves
correspond to: 1) $h^0 V V=g_{h^0 V V}/g_{h_{sm}^0 V V}$, 2) $H^0 V
V=g_{H^0 V V}/g_{h_{sm}^0 V V}$, 3) $|h^0 V V|^2=|g_{h^0 V
V}/g_{h_{sm}^0 V V}|^2$ and 4) $|H^0 V V|^2=|g_{H^0 V V}/g_{h_{sm}^0
V V}|^2$, where $V=W$ or $Z$.} \label{figure15}
\end{center}
\end{figure}

\begin{figure}[floatfix]
\begin{center}
\includegraphics{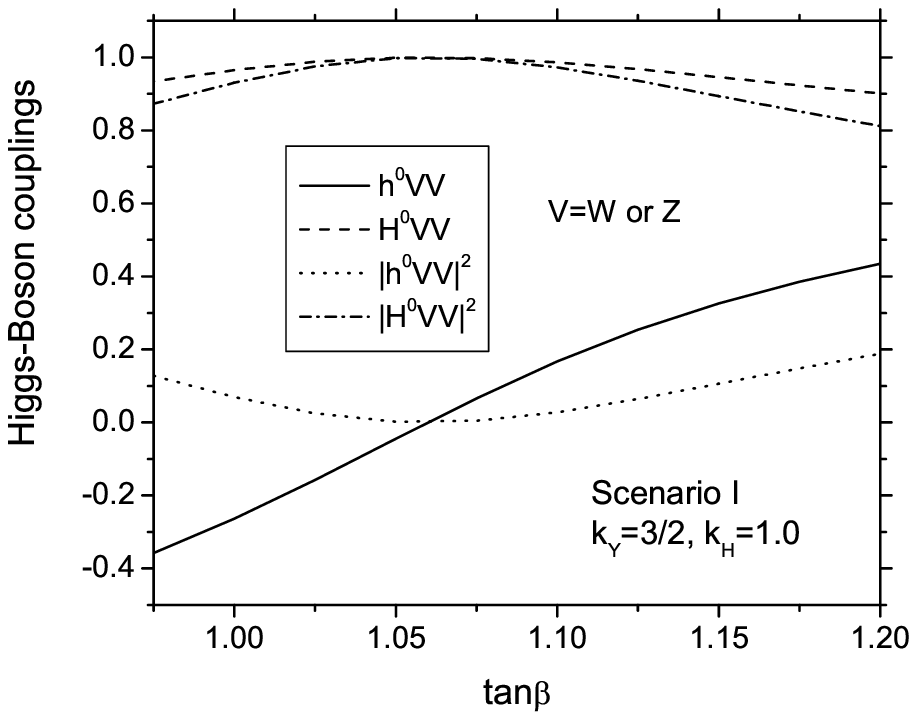}
\caption{Prediction for the Higgs-boson couplings as a function of
$\tan\beta$ in the context of the THDM with $k_Y=3/2$, in the frame
of Scenario I, taking $k_H(1)=1$ and $m_{top}=170.0$ GeV. The curves
correspond to: 1) $h^0 V V=g_{h^0 V V}/g_{h_{sm}^0 V V}$, 2) $H^0 V
V=g_{H^0 V V}/g_{h_{sm}^0 V V}$, 3) $|h^0 V V|^2=|g_{h^0 V
V}/g_{h_{sm}^0 V V}|^2$ and 4) $|H^0 V V|^2=|g_{H^0 V V}/g_{h_{sm}^0
V V}|^2$, where $V=W$ or $Z$.} \label{figure16}
\end{center}
\end{figure}

\begin{figure}[floatfix]
\begin{center}
\includegraphics{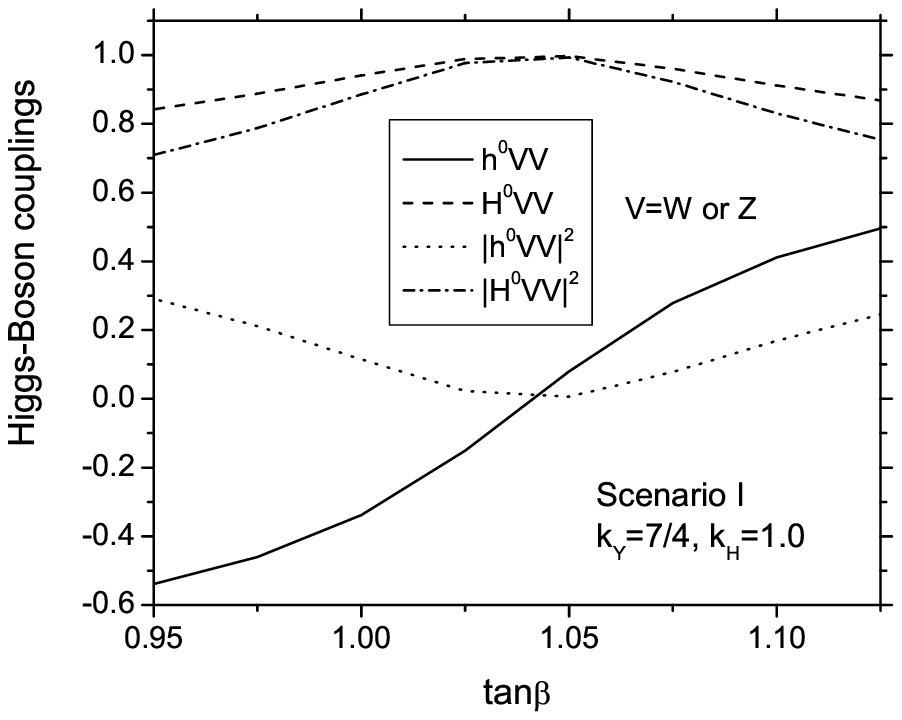}
\caption{Prediction for the Higgs-boson couplings as a function of
$\tan\beta$ in the context of the THDM with $k_Y=7/4$, in the frame
of Scenario I, taking $k_H(1)=1$ and $m_{top}=170.0$ GeV. The curves
correspond to: 1) $h^0 V V=g_{h^0 V V}/g_{h_{sm}^0 V V}$, 2) $H^0 V
V=g_{H^0 V V}/g_{h_{sm}^0 V V}$, 3) $|h^0 V V|^2=|g_{h^0 V
V}/g_{h_{sm}^0 V V}|^2$ and 4) $|H^0 V V|^2=|g_{H^0 V V}/g_{h_{sm}^0
V V}|^2$, where $V=W$ or $Z$.} \label{figure17}
\end{center}
\end{figure}

\section{EW-Higgs unification in the Two-Higgs doublet model.}

Let us now discuss the implications of EW-Higgs unification for the
two-Higgs doublet model (THDM). This model includes two scalar
doublets ($\Phi_1$, $\Phi_2$), and the Higgs potential can be
written as follows \cite{thdmky}:
\begin{eqnarray}
V(\Phi_1,\Phi_2) &=& \mu^2_1 \Phi_1^\dagger \Phi_1
+ \mu^2_2 \Phi_2^\dagger\Phi_2
+ \lambda_1 (\Phi_1^\dagger \Phi_1)^2
+ \lambda_2 (\Phi_2^\dagger \Phi_2)^2
+ \lambda_3 (\Phi_1^\dagger \Phi_1) (\Phi_2^\dagger \Phi_2) \nonumber\\
 & & + \lambda_4 (\Phi_1^\dagger \Phi_2) (\Phi_2^\dagger \Phi_1)
+ \frac{1}{2} \lambda_5 [(\Phi_1^\dagger \Phi_2)^2 +
    (\Phi_2^\dagger \Phi_1)^2] \ .
\end{eqnarray}

\noindent It is clear that by absorbing a phase in the definition of
$\Phi_2$, one can make $\lambda_5$ real and negative, which pushes
all potential CP violating effects into the Yukawa sector:

\begin{equation}
\lambda_5 \leq 0.
\end{equation}

\noindent In order to avoid spontaneous breakdown of the electromagnetic
$U(1)$ \cite{loren1}, the vacuum expectation values must have the following
form:

\begin{equation}
\langle \Phi_1 \rangle = \left(
\begin{array}{c}
0 \\
v_1
\end{array}
\right) \,\,\, , \,\,\, \langle \Phi_2 \rangle = \left(
\begin{array}{c}
0 \\
v_2
\end{array}
\right) \,\,\, ,
\end{equation}

\noindent $v_1^2 + v_2^2 \equiv v^2 = (246 GeV)^2$. This configuration is
indeed a minimum of the tree level potential if the following conditions
are satisfied.
\begin{eqnarray}
\lambda_1 &\geq& 0 \ , \nonumber \\
\lambda_2 &\geq& 0 \ , \nonumber \\
\lambda_4 + \lambda_5 &\leq& 0 \ , \nonumber \\
4\lambda_1 \lambda_2 &\geq& (\lambda_3 + \lambda_4 + \lambda_5)^2 \ .
\end{eqnarray}

The scalar spectrum in this model includes two CP-even states
($h^0,H^0$), one CP-odd ($A^0$) and two charged Higgs bosons
($H^{\pm}$). The tree level expressions for the masses and mixing
angles are given as follows:

\begin{eqnarray}
\tan\beta &=&  \frac{v_2}{v_1} \ , \\
\sin\alpha &=&  -(\mbox{sgn} \, C) \left[\frac{1}{2}
\frac{\sqrt{(A-B)^2+4C^2}-(B-A)}{\sqrt{(A-B)^2+4C^2}} \right]^{1/2} \ , \\
\cos\alpha &=&  \left[\frac{1}{2}
\frac{\sqrt{(A-B)^2+4C^2}+(B-A)}{\sqrt{(A-B)^2+4C^2}} \right]^{1/2} \ , \\
M^2_{H^{\pm}} &=& - \frac{1}{2} (\lambda_4+\lambda_5) v^2 \ , \\
M^2_{A^0} &=& - \lambda_5 v^2 \ , \\
M^2_{H^0,h^0} &=& \frac{1}{2} \left[A+B {\pm} \sqrt{(A-B)^2+4C^2} \right] \ ,
\end{eqnarray}

\noindent where $A=2 \lambda_1 v^2_1$, $B=2 \lambda_2 v^2_2$,
$C=(\lambda_3 + \lambda_4 + \lambda_5) v_1 v_2$.

\noindent The two Higgs doublet models are described by 7
independent parameters which can be taken to be $\alpha$, $\beta$,
$m_{H^{\pm}}$, $m_{H^{0}}$, $m_{h^{0}}$, $m_{A^{0}}$, while the top
quark mass is given as:

\begin{equation}
m_t = g_t v \sin\beta \ .
\end{equation}

\noindent Now, we write the THDM renormalization group equations at
the one loop level involving the gauge coupling constants
$g_{1,2,3}$, the Higgs self-couplings $\lambda_{1,2,3,4,5}$, the
top-quark Yukawa coupling $g_t$, and the parameter $k_Y$, as follows
\cite{thdmky,langa1}:

\begin{eqnarray}
\frac{dg_i}{dt}  &=&  \frac{b_i^{thdm}}{16 \pi^2 } g^3_i \ , \\
\frac{dg_t}{dt}  &=&  \frac{g_t}{16 \pi^2 } \left[ \frac{9}{2} g^2_t -
 \left(\frac{17}{12k_{Y}}g^2_1+ \frac{9}{4}g^2_2+8g^2_3\right)\right] \ , \\
\frac{d\lambda_{1}}{dt}  &=&  \frac{1}{16 \pi^2 }\left[24\lambda_{1}^2 +
2\lambda_{3}^2 + 2\lambda_{3}\lambda_{4} + \lambda_{5}^2 +
\lambda_{4}^2 - 3\lambda_{1}(3g^2_2+\frac{1}{k_{Y}}g^2_1) +
12\lambda_{1}g^2_t \right. \nonumber \\
 & & \left. + \frac{9}{8}g^4_2 + \frac{3}{4k_{Y}}g^2_1g^2_2 +
 \frac{3}{8k^2_{Y}}g^4_1 - 6g^4_t \right] \ , \\
 \frac{d\lambda_{2}}{dt}  &=&  \frac{1}{16 \pi^2 }\left[24\lambda_{2}^2 +
2\lambda_{3}^2 + 2\lambda_{3}\lambda_{4} + \lambda_{5}^2 +
\lambda_{4}^2 - 3\lambda_{2}(3g^2_2+\frac{1}{k_{Y}}g^2_1)
\right. \nonumber \\
 & & \left. + \frac{9}{8}g^4_2 + \frac{3}{4k_{Y}}g^2_1g^2_2 +
 \frac{3}{8k^2_{Y}}g^4_1 \right] \ , \\
\frac{d\lambda_{3}}{dt}  &=&  \frac{1}{16 \pi^2 }\left[
4(\lambda_{1} + \lambda_{2})(3\lambda_{3} + \lambda_{4})
+ 4\lambda^2_{3}+2\lambda^2_{4} + 2\lambda_{5}^2
- 3\lambda_{3}(3g^2_2+\frac{1}{k_{Y}}g^2_1) \right. \nonumber \\
& & \left. + 6\lambda_{3}g^2_t+ \frac{9}{4}g^4_2 - \frac{3}{2k_{Y}}g^2_1g^2_2
+ \frac{3}{4k^2_{Y}}g^4_1 \right] \ , \\
\frac{d\lambda_{4}}{dt}  &=&  \frac{1}{16 \pi^2 }\left[
4\lambda_{4}(\lambda_{1} + \lambda_{2} + 2\lambda_{3} + \lambda_{4})
+ 8\lambda_{5}^2
- 3\lambda_{4}(3g^2_2+\frac{1}{k_{Y}}g^2_1) \right. \nonumber \\
& & \left. + 6\lambda_{4}g^2_t+ \frac{3}{k_{Y}}g^2_1g^2_2
\right] \ , \\
\frac{d\lambda_{5}}{dt}  &=&  \frac{1}{16 \pi^2 }\left[\lambda_{5}
\left((4\lambda_{1} + 4\lambda_{2} + 8\lambda_{3} + 12\lambda_{4} -
3(3g^2_2+\frac{1}{k_{Y}}g^2_1) + 6g^2_t \right) \right] \ ,
\end{eqnarray}
where $(b_1^{thdm},b_2^{thdm},b_3^{thdm})=(7/k_{Y},-3,-7)$; $\mu$
denotes the scale at which the coupling constants are defined, and
$t=\log(\mu/\mu_0)$.

\noindent The form of the unification condition will depend on the
particular realization of this scenario, which could be as generic
as possible. However, in order to make predictions for the Higgs
mass, we shall consider again two specific realizations. Scenario I
will be based on the linear relation:
\begin{equation}
g_1=g_2=k_H(i) \, \lambda_i(M_{GH}) \hspace{2.5cm} (i=1,2,3,4,5) \, ,
\end{equation}
where the factors $k_H(i)$ are included in order to take into
account possible unknown group theoretical or normalization factors.
We shall also define Scenario II, which uses quadratic unification
conditions, as follows:
\begin{equation}
g^2_1=g^2_2=k_H(i) \, \lambda_i(M_{GH}) \hspace{2.5cm} (i=1,2,3,4,5) \, .
\end{equation}

Now, we present first the results of the numerical analysis for the
Higgs bosons masses in the context of the two Higgs-doublet model
for $\tan\beta=1$ and taking $m_{top}=170.0$ GeV. In order to get an
idea of the behavior of the masses of the Higgs bosons ($h^0$,
$H^0$, $A^0$, $H^{\pm}$) we make the following {\it ad hoc} choice:
\begin{equation}
-k_H(5)=\frac{1}{2} k_H(4)=\frac{3}{2} k_H(3)=k_H(2)=k_H(1) \, ,
\end{equation}
for both Scenarios I and II.

It will be also presented in this section a complete discussion on
the resulting couplings of the neutral CP even Higgs bosons with
gauge vector boson pairs in the THDM, which are related to the
corresponding SM couplings as follows \cite{HHG}:
\begin{equation}\label{bosoncoup} \frac{g_{h^0 V
V}}{g_{h_{sm}^0 V V}}=\sin(\beta-\alpha)\, , \mbox{\hspace{1cm}}
\frac{g_{H^0 V V}}{g_{h_{sm}^0 V V}}=\cos(\beta-\alpha)\, ,
\end{equation}
where $V=W$ or $Z$. For the moment it suffices to stress that the
factor $\sin^2(\beta-\alpha)$ fixes the coupling of the lightest CP
even Higgs boson with ZZ pairs, relative to the SM value, and
therefore scales the result for the cross-section of the reaction
$e^+ \, e^- \to h^0 + Z$, which in turn allow us to determine the
Higgs masses within LEP bounds. Hence, results for the Higgs bosons
masses and $\sin^2(\beta-\alpha)$ are given as a function of the
parameter $k_H(1)$, looking for regions which are acceptable
according to the available experimental data. In fact, first we will
make use of the experimental results reported in the Table 14 of
Ref.\cite{unknown:2006cr} which allow, assuming SM decay rates, a
simple and direct check of our results for $m_{h^0}$ and
$\sin^2(\beta-\alpha)$. We would like to emphasize the following:
Even though the analysis of the EW-Higgs Unification within the THDM
implies that the lightest neutral CP-even Higgs boson has a mass
($\sim 100$ GeV) that is somewhat below the LEP bounds, 114.4 GeV
\cite{unknown:2006cr,Barate:2003sz}, it should be mentioned that
this bound refers to the SM Higgs boson. The bound on the lightest
Higgs boson of the THDM depends on the factor
$\sin^2(\beta-\alpha)$, which could be less than 1, thus resulting
in weaker Higgs boson mass bounds. Secondly, we will use the
experimental bound reported for $m_{H^{\pm}}$ in the literature
\cite{Heister:2002ev}:
\begin{equation}\label{bound1}
m_{H^{\pm}} > 79.3 \, \mbox{GeV \hspace{1.0cm} (95\% C.L.)} \, ,\\
\end{equation}
Even though these two comparisons lead to a parameter space
drastically reduced, from Figs.1-8 we observe that there is still an
allowed region for Scenarios I and II, {\it viz}, $0.4 \lsim k_H(1)
\lsim 1.1$ for Scenario I and $0.15 \lsim k_H(1) \lsim 0.55$ for
Scenario II.

From now on, we will restrict ourselves to continue our numerical
analysis only in Scenario I, assuming $k_H(1)=1$

Now we present results in terms of the parameter $\tan\beta$ for the
lightest neutral and the charged Higgs boson masses ($m_{h^0}$ and
$m_{H^{\pm}}$) and the coupling of the lightest neutral CP-even
Higgs boson with $ZZ$ pairs, relative to the corresponding SM value
($|g_{h^0 Z Z}/g_{h_{sm}^0 Z Z}|^2=\sin^2(\beta-\alpha)$), looking
again for regions which are acceptable according to the currently
available experimental data, for $k_Y=5/3$ (Table~\ref{tab:t1}),
$k_Y=3/2$ (Table~\ref{tab:t2}) and $k_Y=7/4$ (Table~\ref{tab:t3}).
As can be seen from Tables~\ref{tab:t1}-\ref{tab:t3} there are
values of $\tan\beta$ where the ratio $|g_{h^0 Z Z}/g_{h_{sm}^0 Z
Z}|^2$ is substantially reduced, which therefore will allow to
overcome the constraints imposed by the LEP search for neutral Higgs
bosons. Lastly, taking into account the bound on $m_{H^{\pm}}$ given
in expression (\ref{bound1}), we conclude that the following regions
for $\tan\beta$ are experimentally allowed:
\begin{equation}\label{range1}
0.975 \leq \tan\beta \leq 1.15 \, \mbox{\hspace{1.0cm} for } k_Y=5/3 \, ,\\
\end{equation}
\begin{equation}\label{range2}
0.975 \leq \tan\beta \leq 1.20 \, \mbox{\hspace{1.0cm} for } k_Y=3/2 \, ,\\
\end{equation}
\begin{equation}\label{range3}
0.95 \leq \tan\beta \leq 1.125 \, \mbox{\hspace{1.0cm} for } k_Y=7/4 \, .\\
\end{equation}

Now, using the ranges given in (\ref{range1}),(\ref{range2}), and
(\ref{range3}) we plot in Figs. 9, 10, and 11, the results for the
Higgs boson masses as a function of $\tan\beta$ for $k_Y=5/3$,
$k_Y=3/2$ and $k_Y=7/4$, respectively. We present also the same
results in Tables \ref{tab:t4}, \ref{tab:t5} and \ref{tab:t6}.

Let us discuss briefly the results of the numerical analysis of the
Higgs mass spectrum. For $k_Y=5/3$, we find that $M_{GH}= 1.3 \times
10^{13}$ GeV, and by taking $\tan\beta=1$ we obtain the following
Higgs mass spectrum $m_{h^0} = 109.1$ GeV, $m_{H^0} = 123.2$ GeV,
$m_{A^0} = 115.5$ GeV, and $m_{H^{\pm}} = 80.3$ GeV. In turn, for
$k_Y=3/2$ we find that $M_{GH}= 5.9 \times 10^{14}$ GeV, somewhat
higher than in the previous case, but for which one still gets a
mass gap between $M_{GH}$ and a possible $M_{GUT}$. One finds
similar values for $m_{H^0}$, $m_{A^0}$, and $m_{H^{\pm}}$ and a
little lower values for $m_{h^0}$, for instance for $\tan\beta=1$,
we get $m_{h^0} = 102.3$ GeV, $m_{H^0} = 122.4$ GeV, $m_{A^0} =
112.8$ GeV, and $m_{H^{\pm}} = 80.3$ GeV. On the other hand, for
$k_Y=7/4$ we find that $M_{GH}= 2.2 \times 10^{12}$ GeV, which is
lower than that of the previous cases, and has an even larger mass
gap between $M_{GH}$ and a possible $M_{GUT}$. We obtain similar
values for $m_{H^0}$, $m_{A^0}$, and $m_{H^{\pm}}$ and a little
higher values for $m_{h^0}$. For instance, for $\tan\beta=1$, one
gets $m_{h^0} = 112.3$ GeV, $m_{H^0} = 123.0$ GeV, $m_{A^0} = 117.4$
GeV, and $m_{H^{\pm}} = 80.6$ GeV.

The numerical results presented in Figs. 9-11 (Tables
\ref{tab:t4}-\ref{tab:t6}) lead us to conclude that the Higgs mass
spectrum is almost independent of the value of $k_Y$. However, the
unification scale $M_{GH}$ depends strongly on the value of $k_Y$,
going from $2.2 \times 10^{12}$ GeV up to $5.9 \times 10^{14}$ GeV
for $7/4 > k_Y > 3/2$ (for $k_Y=5/3$, it is obtained $M_{GH}=1.3
\times 10^{13}$ GeV).

At this point, we want to recall the relation between the
Higgs-fermion couplings, which can be expressed relative to the SM
value and is given by \cite{HHG}:
\begin{eqnarray}\label{fermioncoup}
H^0 t \bar{t} : \frac{\sin\alpha}{\sin\beta} \, ,& \mbox{\hspace{2cm}} &H^0 b \bar{b}:\frac{\cos\alpha}{\cos\beta}  \, , \nonumber\\
h^0 t \bar{t} : \frac{\cos\alpha}{\sin\beta} \, ,&  &h^0 b \bar{b}:\frac{-\sin\alpha}{\cos\beta}  \, , \\
A^0 t \bar{t} : \cot\beta \, ,&  &A^0 b \bar{b}:\tan\beta \, .
\nonumber
\end{eqnarray}

Now, we use the ranges given in (\ref{range1}),(\ref{range2}), and
(\ref{range3}), and we present in Figs. 12, 13, and 14, the results
for the fermion couplings a a function of $\tan\beta$ for $k_Y=5/3$,
$k_Y=3/2$ and $k_Y=7/4$, respectively.

Finally, making use of the ranges given in
(\ref{range1}),(\ref{range2}), and (\ref{range3}), we present in
Figs. 15, 16, and 17, the results for the Higgs-boson couplings as a
function of $\tan\beta$ for $k_Y=5/3$, $k_Y=3/2$ and $k_Y=7/4$,
respectively.

From our results shown in Figs. 9-17 and Tables
\ref{tab:t4}-\ref{tab:t6}, we also conclude that the Higgs mass
spectrum does not depend strongly on the value of $\tan\beta$. On
the other hand, the fermion couplings and the boson couplings depend
strongly on the value of $\tan\beta$.

We find that for $\tan\beta=1$, the coupling of $h^0$ to up-type
(d-type) quarks is suppressed (enhanced), which will have important
phenomenological consequences \cite{Balazs:1998nt,Diaz-Cruz:1998qc}:
For instance it will suppress the production of $h^0$ at hadron
colliders through gluon fusion, while the associated production with
$b\bar{b}$ quarks will be enhanced. The couplings of $H^0$ show the
opposite behavior, namely the couplings with d-type (up-type) quarks
are suppressed (enhanced). This behavior changes as $\tan\beta$
takes higher values, and is reversed already for $\tan\beta=1.1$.
Similar results are obtained for other normalizations. We end this
section by saying that similar results are obtained in the
experimentally allowed regions for Scenario II.

\begin{table}
\begin{tabular}{|c|c|c|c|}
\hline

 $\;\tan\beta\;$ & $\;m_{h^0}\;$ (GeV) & $\;\sin^{2}(\beta-\alpha)\;$ & $\;m_{H^{\pm}}$ (GeV)$\;$ \\

\hline \hline

$ \;0.900 \;$&$\; 105.3 \;$&$\; 0.3674 \;$&$\; 77.26\;$\\
$ \;0.925 \;$&$\; 106.5 \;$&$\; 0.3155 \;$&$\; 78.13\;$\\
$ \;0.950 \;$&$\; 107.6 \;$&$\; 0.2540 \;$&$\; 78.92\;$\\
$ \;0.975 \;$&$\; 108.5 \;$&$\; 0.1816 \;$&$\; 79.63\;$\\
$ \;1.000 \;$&$\; 109.1 \;$&$\; 0.1020 \;$&$\; 80.28\;$\\
$ \;1.025 \;$&$\; 109.3 \;$&$\; 0.0315 \;$&$\; 80.87\;$\\
$ \;1.050 \;$&$\; 109.0 \;$&$\; 0.0001 \;$&$\; 81.41\;$\\
$ \;1.075 \;$&$\; 108.2 \;$&$\; 0.0238 \;$&$\; 81.90\;$\\
$ \;1.100 \;$&$\; 107.0 \;$&$\; 0.0828 \;$&$\; 82.36\;$\\
$ \;1.125 \;$&$\; 105.6 \;$&$\; 0.1497 \;$&$\; 82.78\;$\\
$ \;1.150 \;$&$\; 104.0 \;$&$\; 0.2107 \;$&$\; 83.17\;$\\
$ \;1.175 \;$&$\; 102.4 \;$&$\; 0.2629 \;$&$\; 83.53\;$\\
$ \;1.200 \;$&$\; 100.8 \;$&$\; 0.3068 \;$&$\; 83.86\;$\\
$ \;1.225 \;$&$\; 99.24 \;$&$\; 0.3442 \;$&$\; 84.17\;$\\
$ \;1.250 \;$&$\; 97.70 \;$&$\; 0.3763 \;$&$\; 84.46\;$\\
$ \;1.275 \;$&$\; 96.21 \;$&$\; 0.4044 \;$&$\; 84.73\;$\\
$ \;1.300 \;$&$\; 94.75 \;$&$\; 0.4293 \;$&$\; 84.99\;$\\

\hline
\end{tabular}
\caption{Prediction for the lightest neutral Higgs boson $h^0$ mass,
$\sin^{2}(\beta-\alpha)$ and the charged Higgs boson $H^{\pm}$ mass
as a function of $\tan\beta$ in the context of the THDM with
$k_Y=5/3$, in the frame of Scenario I with $k_H(1)=1$, taking
$m_{top}=170.0$ GeV.}\label{tab:t1}
\end{table}

\newpage

\begin{table}
\begin{tabular}{|c|c|c|c|}
\hline

 $\;\tan\beta\;$ & $\;m_{h^0}\;$ (GeV) & $\;\sin^{2}(\beta-\alpha)\;$ & $\;m_{H^{\pm}}$ (GeV)$\;$ \\

\hline \hline

 $ \;0.900 \;$&$\; 99.93 \;$&$\; 0.3063 \;$&$\; 76.91\;$\\
 $ \;0.925 \;$&$\; 100.8 \;$&$\; 0.2503 \;$&$\; 77.89\;$\\
 $ \;0.950 \;$&$\; 101.6 \;$&$\; 0.1898 \;$&$\; 78.78\;$\\
 $ \;0.975 \;$&$\; 102.1 \;$&$\; 0.1277 \;$&$\; 79.57\;$\\
 $ \;1.000 \;$&$\; 102.3 \;$&$\; 0.0697 \;$&$\; 80.30\;$\\
 $ \;1.025 \;$&$\; 102.3 \;$&$\; 0.0249 \;$&$\; 80.95\;$\\
 $ \;1.050 \;$&$\; 101.9 \;$&$\; 0.0020 \;$&$\; 81.55\;$\\
 $ \;1.075 \;$&$\; 101.2 \;$&$\; 0.0044 \;$&$\; 82.10\;$\\
 $ \;1.100 \;$&$\; 100.2 \;$&$\; 0.0279 \;$&$\; 82.61\;$\\
 $ \;1.125 \;$&$\; 99.06 \;$&$\; 0.0643 \;$&$\; 83.07\;$\\
 $ \;1.150 \;$&$\; 97.78 \;$&$\; 0.1060 \;$&$\; 83.50\;$\\
 $ \;1.175 \;$&$\; 96.43 \;$&$\; 0.1483 \;$&$\; 83.90\;$\\
 $ \;1.200 \;$&$\; 95.04 \;$&$\; 0.1886 \;$&$\; 84.27\;$\\
 $ \;1.225 \;$&$\; 93.65 \;$&$\; 0.2259 \;$&$\; 84.61\;$\\
 $ \;1.250 \;$&$\; 92.27 \;$&$\; 0.2601 \;$&$\; 84.93\;$\\
 $ \;1.275 \;$&$\; 90.90 \;$&$\; 0.2913 \;$&$\; 85.23\;$\\
 $ \;1.300 \;$&$\; 89.56 \;$&$\; 0.3198 \;$&$\; 85.51\;$\\

\hline
\end{tabular}
\caption{Prediction for the lightest neutral Higgs boson $h^0$ mass,
$\sin^{2}(\beta-\alpha)$ and the charged Higgs boson $H^{\pm}$ mass
as a function of $\tan\beta$ in the context of the THDM with
$k_Y=3/2$, in the frame of Scenario I with $k_H(1)=1$, taking
$m_{top}=170.0$ GeV.}\label{tab:t2}
\end{table}

\newpage

\begin{table}
\begin{tabular}{|c|c|c|c|}
\hline

 $\;\tan\beta\;$ & $\;m_{h^0}\;$ (GeV) & $\;\sin^{2}(\beta-\alpha)\;$ & $\;m_{H^{\pm}}$ (GeV)$\;$ \\

\hline \hline

$ \; 0.900 \;$&$\; 107.9 \;$&$\; 0.4004 \;$&$\; 77.79\;$\\
$ \; 0.925 \;$&$\; 109.3 \;$&$\; 0.3518 \;$&$\; 78.61\;$\\
$ \; 0.950 \;$&$\; 110.5 \;$&$\; 0.2908 \;$&$\; 79.35\;$\\
$ \; 0.975 \;$&$\; 111.6 \;$&$\; 0.2121 \;$&$\; 80.02\;$\\
$ \; 1.000 \;$&$\; 112.3 \;$&$\; 0.1144 \;$&$\; 80.63\;$\\
$ \; 1.025 \;$&$\; 112.6 \;$&$\; 0.0229 \;$&$\; 81.18\;$\\
$ \; 1.050 \;$&$\; 112.2 \;$&$\; 0.0064 \;$&$\; 81.69\;$\\
$ \; 1.075 \;$&$\; 111.1 \;$&$\; 0.0777 \;$&$\; 82.16\;$\\
$ \; 1.100 \;$&$\; 109.5 \;$&$\; 0.1693 \;$&$\; 82.58\;$\\
$ \; 1.125 \;$&$\; 107.9 \;$&$\; 0.2460 \;$&$\; 82.98\;$\\
$ \; 1.150 \;$&$\; 106.1 \;$&$\; 0.3054 \;$&$\; 83.34\;$\\
$ \; 1.175 \;$&$\; 104.4 \;$&$\; 0.3519 \;$&$\; 83.68\;$\\
$ \; 1.200 \;$&$\; 102.7 \;$&$\; 0.3893 \;$&$\; 84.00\;$\\
$ \; 1.225 \;$&$\; 101.1 \;$&$\; 0.4204 \;$&$\; 84.29\;$\\
$ \; 1.250 \;$&$\; 99.50 \;$&$\; 0.4470 \;$&$\; 84.56\;$\\
$ \; 1.275 \;$&$\; 97.97 \;$&$\; 0.4701 \;$&$\; 84.82\;$\\
$ \; 1.300 \;$&$\; 96.48 \;$&$\; 0.4907 \;$&$\; 85.06\;$\\

\hline
\end{tabular}
\caption{Prediction for the lightest neutral Higgs boson $h^0$ mass,
$\sin^{2}(\beta-\alpha)$ and the charged Higgs boson $H^{\pm}$ mass
as a function of $\tan\beta$ in the context of the THDM with
$k_Y=7/4$, in the frame of Scenario I with $k_H(1)=1$, taking
$m_{top}=170.0$ GeV.}\label{tab:t3}
\end{table}

\newpage

\begin{table}
\begin{tabular}{|c|c|c|c|c|}
\hline

 $\;\tan\beta\;$ & $\;m_{h^0}\;$ (GeV) & $\;m_{H^0}\;$ (GeV)  & $\;m_{A^0}\;$ (GeV)  & $\;m_{H^{\pm}}$ (GeV)$\;$ \\

\hline \hline

 $\;0.975\;$&$\;108.5\;$&$\;125.4\;$&$\;114.7\;$&$\;79.63\;$\\
 $\;1.000\;$&$\;109.1\;$&$\;123.2\;$&$\;115.5\;$&$\;80.28\;$\\
 $\;1.025\;$&$\;109.3\;$&$\;121.6\;$&$\;116.2\;$&$\;80.87\;$\\
 $\;1.050\;$&$\;109.0\;$&$\;120.6\;$&$\;116.9\;$&$\;81.41\;$\\
 $\;1.075\;$&$\;108.2\;$&$\;120.2\;$&$\;117.6\;$&$\;81.90\;$\\
 $\;1.100\;$&$\;107.0\;$&$\;120.3\;$&$\;118.2\;$&$\;82.36\;$\\
 $\;1.125\;$&$\;105.6\;$&$\;120.7\;$&$\;118.7\;$&$\;82.78\;$\\
 $\;1.150\;$&$\;104.0\;$&$\;121.3\;$&$\;119.2\;$&$\;83.17\;$\\

\hline
\end{tabular}
\caption{Predicted Higgs mass spectrum as a function of $\tan\beta$
in the context of the THDM with $k_Y=5/3$, in the frame of Scenario
I with $k_H(1)=1$, taking $m_{top}=170.0$ GeV.}\label{tab:t4}
\end{table}

\begin{table}
\begin{tabular}{|c|c|c|c|c|}
\hline

 $\;\tan\beta\;$ & $\;m_{h^0}\;$ (GeV) & $\;m_{H^0}\;$ (GeV)  & $\;m_{A^0}\;$ (GeV)  & $\;m_{H^{\pm}}$ (GeV)$\;$ \\

\hline \hline

 $\;0.975\;$&$\;102.1\;$&$\;124.4\;$&$\;111.8\;$&$\;79.57\;$\\
 $\;1.000\;$&$\;102.3\;$&$\;122.4\;$&$\;112.8\;$&$\;80.30\;$\\
 $\;1.025\;$&$\;102.3\;$&$\;120.9\;$&$\;113.6\;$&$\;80.95\;$\\
 $\;1.050\;$&$\;101.9\;$&$\;119.9\;$&$\;114.3\;$&$\;81.55\;$\\
 $\;1.075\;$&$\;101.2\;$&$\;119.3\;$&$\;115.0\;$&$\;82.10\;$\\
 $\;1.100\;$&$\;100.2\;$&$\;119.1\;$&$\;115.6\;$&$\;82.61\;$\\
 $\;1.125\;$&$\;99.06\;$&$\;119.1\;$&$\;116.2\;$&$\;83.07\;$\\
 $\;1.150\;$&$\;97.78\;$&$\;119.4\;$&$\;116.8\;$&$\;83.50\;$\\
 $\;1.175\;$&$\;96.43\;$&$\;119.7\;$&$\;117.3\;$&$\;83.90\;$\\
 $\;1.200\;$&$\;95.04\;$&$\;120.2\;$&$\;117.7\;$&$\;84.27\;$\\

\hline
\end{tabular}
\caption{Predicted Higgs mass spectrum as a function of $\tan\beta$
in the context of the THDM with $k_Y=3/2$, in the frame of Scenario
I with $k_H(1)=1$, taking $m_{top}=170.0$ GeV.}\label{tab:t5}
\end{table}

\begin{table}
\begin{tabular}{|c|c|c|c|c|}
\hline

 $\;\tan\beta\;$ & $\;m_{h^0}\;$ (GeV) & $\;m_{H^0}\;$ (GeV)  & $\;m_{A^0}\;$ (GeV)  & $\;m_{H^{\pm}}$ (GeV)$\;$ \\

\hline \hline

 $\;0.950\;$&$\;110.5\;$&$\;127.9\;$&$\;115.7\;$&$\;79.35\;$\\
 $\;0.975\;$&$\;111.6\;$&$\;125.2\;$&$\;116.6\;$&$\;80.02\;$\\
 $\;1.000\;$&$\;112.3\;$&$\;123.0\;$&$\;117.4\;$&$\;80.63\;$\\
 $\;1.025\;$&$\;112.6\;$&$\;121.3\;$&$\;118.1\;$&$\;81.18\;$\\
 $\;1.050\;$&$\;112.2\;$&$\;120.6\;$&$\;118.8\;$&$\;81.69\;$\\
 $\;1.075\;$&$\;111.1\;$&$\;120.6\;$&$\;119.4\;$&$\;82.16\;$\\
 $\;1.100\;$&$\;109.5\;$&$\;121.1\;$&$\;119.9\;$&$\;82.58\;$\\
 $\;1.125\;$&$\;107.9\;$&$\;121.8\;$&$\;120.5\;$&$\;82.98\;$\\

\hline
\end{tabular}
\caption{Predicted Higgs mass spectrum as a function of $\tan\beta$
in the context of the THDM with $k_Y=7/4$, in the frame of Scenario
I with $k_H(1)=1$, taking $m_{top}=170.0$ GeV.}\label{tab:t6}
\end{table}

\section{Comments and conclusions}

In this paper we have obtained the Higgs mass spectrum, the
Higgs-fermion couplings and the Higgs-boson couplings of the THDM in
a framework where it is possible to unify the Higgs self-coupling
with the gauge interactions.

The hypercharge normalization plays an important role to identify
the EW-Higgs unification scale. For the canonical value $k_Y=5/3$ we
get $M_{GH} = 1.3 \times 10^{13}$ GeV. For lower values, such as
$k_Y=3/2$ the scale is $M_{GH} = 5.9 \times 10^{14}$ GeV, which is
closer to the GUT scale ($\approx 10^{16}$ GeV) but for higher
values, such as $k_Y=7/4$, which gives $M_{GH} = 2.2 \times 10^{12}$
GeV, the EW-Higgs unification becomes clearly distinctive.

The present approach still lacks a solution to the hierarchy
problem; at the moment we have to affiliate to argument that
fundamental physics could accept some fine-tuning \cite{splitsusy}.
Another option would be to consider one of the simplest early
attempts to solve the problem of quadratic divergences in the SM,
namely through an accidental cancellation~\cite{veltmanqd}. In fact,
such kind of cancellation implies a relationship between the quartic
Higgs coupling and the Yukawa and gauge constants, which has the
form:
$\lambda = y^2_t -\frac{1}{8} [ 3g^2 +g'^2]$.
Unfortunately, this relation implies a Higgs mass $m_\phi=316$ GeV,
and that seems already excluded. Nevertheless, this relation could
work if one takes into account the running of the coupling and
Yukawa constants. This particular point will be the subject of
future investigations.

\acknowledgments{The authors would like to thank CONACYT and SNI
(Mexico) for financial support and the Huejotzingo Seminar for
inspiring discussions. J.L. Diaz-Cruz also thanks A. Aranda for
interesting discussions.}


\end{document}